\documentstyle[psfig,twocolumn,aps]{revtex}
\begin{document}
\wideabs{
\title{
The upper critical field problem in MgB$_2$.}
\author{S.V.\ Shulga}
 \address{ Institute of Spectroscopy,  RAS, Troitsk, 142190, Russia}
\author{ S.-L.\ Drechsler, H.\ Eschrig}
 \address{ Institut f\"ur Festk\"orper- und Werkstofforschung Dresden e.V.,
 Postfach 270016, D-01171 Dresden, Germany}
\author{H.\ Rosner and W.\ Pickett}
\address{Department of Physics, University of California, Davis, CA 95616, 
USA}
\date{\today}
\maketitle
\begin{abstract}
The upper critical field $H_{c2}(T)$ for
MgB$_{2}$ is analyzed in terms of single and multi-band Eliashberg 
models. The relatively high value of $H_{c2}(0)\approx$ 14 to 18 T 
can be understood,
 if  a strongly 
coupled subgroup of heavy quasiparticles is involved 
in the superconductivity.
The sizable coupling to both low- and high-frequency bosons is essential. 
 This picture is supported by LDA calculations of
  Fermi velocity distribution  over different sheets
 of the Fermi surface with special emphasis on the hole tubes.
The possible origin of anharmonic soft modes  is briefly discussed.
Similarities and differences with transition metal borocarbides
are figured out.
\end{abstract}
}
\narrowtext

 The recent  
 discovery of superconductivity in MgB$_2$\cite{akimitsu00} 
  has initiated  an immediate broad research activity due to the  
 high transition temperature $T_c\sim $ 40 K 
 in a seemingly ordinary $s-p$  metal. 
The present discussions mainly focus on measured quantities
 $T_c$ and  the gap
$2\Delta_0$ with respect to the topics: (1) weak or strong coupling
in terms of standard BCS theory, (2) its possible  
relationship to  the  
superconductivity in transition metal borocarbides, (3)
symmetry of the order parameter,
 and (4) electron-phonon (el-ph)
interaction or Coulomb repulsion based mechanisms 
of pairing  \cite{hirsch,imada}.  
In context of the standard phonon mechanism: 
 (i) a dominant role of intermediately coupled 
 high-frequency boron phonons is assumed  (the simple  
``metallic hydrogen  scenario''); or (ii) a strong
coupling scenario is discussed.  
In all those discussions,
essentially the validity of the standard isotropic single band (ISB)
picture is implicitly assumed.
However, the low temperature value of the 
upper critical field $H_{c2}(0)$, the fundamental 
quantity of a type-II superconductor, has not analyzed so far.  
 In this context we would like to point out 
an interesting
puzzle: with the computed and commonly accepted average band 
Fermi velocity $v_F$=8.9$\times 10^7
$ cm/s and the plasma frequency $\omega_{pl}$=7 eV, the 
resulting 
clean-limit London
penetration depth $\lambda_L=c/\omega_{pl}$=29 nm nearly {\it
equals} the BCS coherence length $\xi_0=0.18\hbar v_F/k_BT_c$=29 nm 
while experiments on $H_c$ and $H_{c2}$ yield a 
 Ginzburg - Landau
parameter $\kappa$=26 \cite{petrovic}.

The ISB model \cite{carbotte90}
is the most developed {\it part} of the  theory of superconductivity. It   
describes {\it quantitatively} the renormalization of
the physical properties of metals  due to the  
electron-phonon interaction. 
The  input material   parameters
are 
the density of states at $E_F$, $N(0)$, the Fermi velocity
$v_F$, the  impurity scattering rate $\gamma_{imp}$,
 the  paramagnetic impurity scattering rate $\gamma_{m}$, the Coulomb 
pseudopotential
$\mu^{*}(\approx 0.1)$, and the electron-phonon spectral function
 $\alpha^2F(\Omega)$. 
The temperature dependencies of physical 
properties are  determined mainly 
by the  first inverse moments of the spectral function - the coupling constant
 $\lambda$=2$\int d\Omega \alpha^2F(\Omega)/ \Omega$
and  the average boson energy
$\Omega_{log}=\exp{[(2/\lambda)\int_0^{\infty}d\Omega
\mbox{\ } log(\Omega)\alpha^2F(\Omega)/\Omega]}$.

Our analysis is based on  numerical calculations of $H_{c2}(T)$
solving  equations first presented by Prohammer {\it et al.} \cite{prohammer87}. 
In the intermediate
and weak coupling cases,  the 
clean limit value of the calculated orbital
$H_{c2}(0)$
practically does not depend on the shape of the spectral function
\cite{carbotte90,shulga98}. 
In the spirit of Langmann \cite{langmann91}
the $s$-wave clean limit ISB-$H_{c2}(0)$ can be approximated by the following
 factorized formula
\begin{equation}
 H_{c2}(0)[{\rm T}]\approx 0.02T_c^2[{\rm K}](1+\lambda)^{2.4}
 /v_F^2[{\rm 10^7 cm/s}] .
\label{hot0}
\end{equation}
It is convenient to express $v_F^2$ in Eq.\ \ref{hot0} using 
the  related experimentally available quantities:
the cell volume $V$, the Sommerfeld constant (per mole)
$\gamma_S=\pi^2k_B^2N(0)(1+\lambda)N_AV/3$ and 
the bare plasma frequency $\omega_{pl}^2=4\pi e^2N(0)v_F^2/3$
\begin{equation}
v_F^2=\frac{\pi N_AV k_B^2(1+\lambda)\omega_{pl}^2}{4e^2\gamma_S}.
\label{hot1}
\end{equation}
 Substituting Eq.\ \ref{hot1}  into  Eq.\ \ref{hot0} one arrives at 
 the relevant ratio
\begin{equation}
Q=\frac{3.6\omega_{pl}^2\mbox{\ }[{\rm eV^2}] \mbox{\ } H_{c2}(0)[{\rm T}]
V[{\rm \AA^3}]}
{\gamma_S {\rm [mJ/mol\times K^2]}T_c^2{\rm [K]}(1+\lambda)^{1.4}},
\label{isbcre}
\end{equation}
the value of which must be $Q\approx$1 in a  ISB-like case.
A  strong deviation   $Q$ from unity indicates the 
inapplicability of the $s$-wave ISB model. 

To proceed within the $Q$-check,  let us estimate the coupling constant
$\lambda$ and the corresponding averaged boson frequency
$\langle\Omega\rangle$.  Recent   neutron data
\cite{taku} show  phonon density of states (PDOS) extending to 
above 100 meV with main peaks at about 90, 55, 37, and 
32 meV,  as well as an unexpected
anomalous soft mode near 17.5 meV 
below the maxima of acoustic modes. 
 The measured PDOS is in agreement
with $\Theta_D$=65 meV, estimated from specific heat
data\cite{canfield}.  The values for zone-center optical mode obtained 
from density-functional theory based calculations
58 meV \cite{pickett}, 62 meV \cite{kortus}, 73 meV \cite{dolgov} are
rather similar and for our goal we adopt an Einstein spectrum with
$\Omega$=60 meV.  With the standard value $\mu^*$=0.1 at the cutoff
frequency of 6$\langle\Omega\rangle$ one needs $\lambda$=0.86
to obtain $T_c$=40 K. If the dominating 
coupling were that by the 17.5 meV feature 
 $\lambda$=2.5 would be needed. 
  
Now we are in a position to apply our  $Q$-test to selected 
superconductors. The results are presented in Table~1 \ref{tab1}.
For the weakly anisotropic transition metal Nb it is approximately fulfilled, 
whereas
strong deviations do occur both for the 
transition metal borocarbide YNi$_2$B$_2$C
 and
for MgB$_2$ under consideration. The latter one
is checked in the weak and as well as in the strong coupling cases.

These difficulties can be resolved in the framework of a more complex
effective $N$-band model.  For an anisotropic system it is obtained by
dividing the Fermi surface into $N$ parts and approximating the {\bf
k}-dependent quantities in each part by their mean values
\cite{langmann92}.  The input material parameters of such a model are
the densities of states at $E_F$ $N_i(0)$, the Fermi velocities
$v_{F,i}$, the impurity scattering rates $\gamma_{imp,ij}$, the
Coulomb pseudopotentials $\mu^{*}_{i}(\approx 0.1)$, and the
electron-phonon spectral functions $\alpha^2_{ij}F(\omega)$.  For
the sake of simplicity, as a first step, we consider here such 
effects in the frame
of an effective two-band model. Recently it  was successfully applied to 
transition metal borocarbides in a combined study of Eliashberg theory 
and de Haas van Alphen effect\cite{shulga98}. Unfortunately, 
due to the lack of 
single crystals, at present there 
is no available experimental information on the electronic structure near the 
Fermi surface of MgB$_2$.
For this reason, we performed electronic structure
calculations on a very dense $\vec{k}$-mesh
 using the full-potential nonorthogonal
local-orbital minimum-basis scheme \cite{koepernik99}.
Thus we get more detailed  information on the Fermi velocities 
compared with 
published data \cite{kortus}. 

\begin{figure}
\psfig{figure=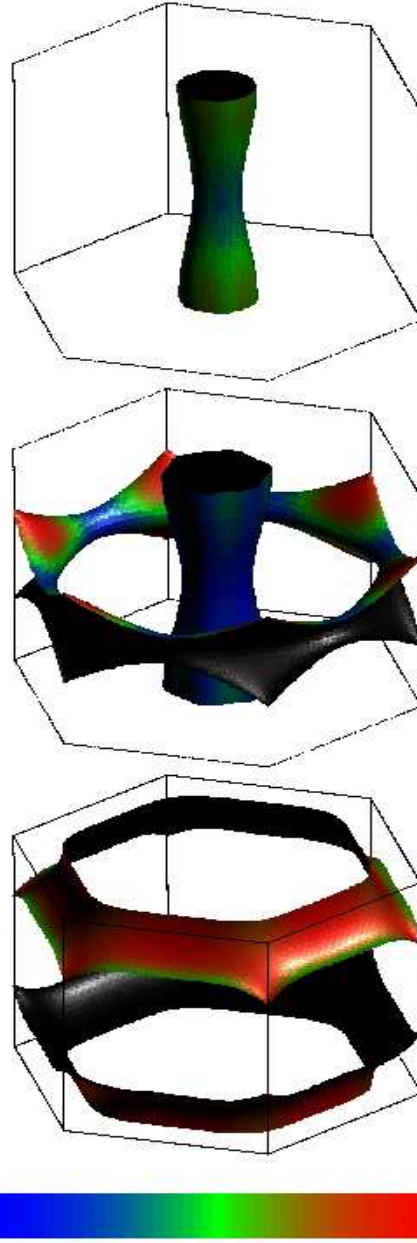,width=7.0cm,height=17.cm}
\caption{
 (Color) The Fermi surface of MgB$_2$ in the full Brillouin-zone. 
The $\Gamma$-point is  the midpoint of the hexagonal prisma 
and the A-point(s) are  the midpoints of the hexagons at the 
top and the bottom of the prisma,
respectively.
Light-hole tube (upper panel);
Heavy-hole tube inside the hole honeycomb (middle panel);
Electron honeycomb (lower panel).
The Fermi velocities are measured in units of 10$^7$ cm/s 
presented by different colors shown at the bottom of the figure.} 
\label{fig1}
\end{figure}

The Fermi surface of MgB$_2$ (see Fig.\ \ref{fig1}) consists of four
disjoint sheets: two hexagonal heavy hole tubes around the $\Gamma -A$
line and a light hole and an electron honeycombs.  In the spirit of a
two-band model we divide the quasiparticles into subgroups with
approximately similar material parameters.  We choose the heavy-hole
tube as the first effective band (central blue piece in the middle of
Fig.\ \ref{fig1}). Its DOS contribution $N_h(0)\equiv N_1(0)$ is about
a third of the total DOS: $N(0)/N_1(0)\approx$3 and we adopt
$N_2(0)/N_1(0)$=2. The average Fermi velocity over this hole tube is
$v_{F1}$=4.5$\times$10$^7$ cm/s while for the remaining fast Fermi surface sheets we
choose $v_{F2}$=11$\times$10$^7$ cm/s. In the clean limit the
scattering rates $\gamma_{imp,ij}$ can be set to 0.  Lets start with
weak-coupling scenario.  If one could ignore the fast quasiparticles
entirely, one would return to the ISB case, considered above, with
$\lambda$ = 0.8 and $v_F$=4.5$\times$10$^7$.  From Eq.1 one then gets
$H_{c2}(0)$=6.5 T which is still much smaller than the experimental
value. Moreover, within a consequent isotropic coupling scenario the
interband coupling cannot be ignored.  As a result, we arrive at 2T,
only (see Fig.\ 2). For comparison the ISB case for the same $\lambda$
and $v_F$=8.9 $\times$ 10$^7$cm/s is shown, too.  For this moderate
anisotropy of $v_F$, the two-band effect is practically unimportant.
Hence, to describe $H_{c2}(T)$ quantitatively, an enhanced coupling of
the heavy holes would be needed.  Deformation potential calculations
provide support for such an enhancement for the larger heavy-hole tube
\cite{pickett}.
  
It was shown above, that at  given 
$T_c=40$ K and el-ph interaction 
with high-frequency phonons, only, the coupling constant cannot exceed 
unity and $H_{c2}(0)$ doesn't exceed several Tesla at most. To increase
the coupling constant, we are forced to adopt strong 
coupling to low-frequency mode(s).
Hence,  
 we adopt a two-peak spectral function containing for instance 
a low-frequency peak at 17.5 meV as well as an effective 
high-frequency peak at 60 meV with comparable spectral weights.
We adopt $\lambda_{60 meV}/\lambda_{17.5 meV}$=0.33. 
Then reasonable agreement with available experimental data 
for $H_{c2}(T)$ \cite{budko,fuchs}
can be achieved  with $\lambda_{11}$=1.5,  $\lambda_{22}$=0.4
and  $\lambda_{21}$=0.5 (see Fig.2).  With the bare band value 
N(0)=1.7 mJ/mol K$^2$ \cite{pickett} 
we arrive at $\gamma_S$=4.1 mJ/mol K$^2$ and
$\omega_{pl}$=7.5 eV in accord with experimental data 
\cite{canfield,dolgov,ott}. The calculated gap values  $2\Delta_1$=5$k_BT_c$
and $2\Delta_2$=3$k_BT_c$ above and below the BCS value 3.5$k_BT_c$,
 respectively, can
 be compared with the experimental
 values taken from
tunneling  (3.4$k_BT_c$, 4.83$k_BT_c$)  \cite{kara,schmidt}
and specific heat (2.4$k_BT_c$) \cite{ott}  and NMR 
data (2.5$k_BT_c$,5.$k_BT_c$) 
\cite{kote,gerashenko}.

It is interesting to realize, that  switching off the soft mode at fixed 
remaining 
parameters results in $T_{c,hard} \approx$ 7K, only. However, switching off 
instead the 
high-frequency modes, we would arrive at $T_{c,soft} \approx$ 20K at least.
Alternatively, if the spectral functions differ 
significantly for various Fermi surface 
 sheets, the contribution of the high-frequency modes to $T_c$ can 
enhanced up to about 20 K.
Anyhow, this means, that the ``medium'' high $T_c$ superconductivity 
 of MgB$_2$ within our model is not provided by a simple 
 ``hydrogen-like'' scenario
as frequently discussed in the literature,
but by a more complex scenario, where both $T_c$ and $H_{c2}$(0) {\it do} 
depend very sensitively on the coupling with the soft-mode.
A somewhat similar interplay of high- and low-frequency phonons occurs 
in the well-known PdH(D) palladium hydrides (deuterides).
 
The strong coupling in the heavy-hole band
due to interaction with the soft mode needs some interpretation and discussion. 
A simple down-tuning of the calculated $E_{2g}$-mode  
frequency from 58.5 meV \cite{kortus} to 17.5 meV, preserving thereby 
the harmonic 
approximation, would result in large zero-point fluctuations amplitudes
 for which there
is not enough space in a real crystal. In addition we would arrive at  
an unrealistic super strong coupling limit:
$\lambda_1$=10.4 adopting the estimated Hopfield parameter
$\eta_1=$ 8.2eV/\AA$^2$ \cite{pickett}.
These difficulties can be resolved adopting
an anharmonic model for the lattice vibrations.
Within its  extreme anharmonic two-level system
approximation
 (deep double- or multiple wells)   
\cite{plakida,galbaatar} the corresponding anharmonic electron-boson 
coupling constant reads 
\begin{equation}
\lambda_1(T)=\frac{2}{3}\eta_1d_0^2\tanh 
(\hbar \omega_1/2k_BT)/\hbar\omega_1,
\label{anharmlambda}
\end{equation}
 where $d_0\sim 0.1$ \AA \ is a typical double-well distance. It
is sizable and at the same time not too large:
thus we arrive at $\lambda_1$=2 in accord with $\lambda_{11}+\lambda_{12}$ 
obtained above phenomenologically. Quite interestingly, in this limit the
boson frequency remains constant whereas the intensity of its spectral density 
increases with
decreasing $T$ like in 
the experimental neutron data\cite{taku}
 until saturation at low $T$ is 
achieved. For the adopted value of $\omega_1$=17.5 meV this occurs below 50 K,
 i.e.\  only slightly above $T_c$. 
Alternatively, in principle, a softening of other modes 
 than the $E_{2g}$ mode
should be considered, too. 
Guided by the experience obtained from the somewhat related 
transition metal borocarbides (see below), a softening of the acoustic modes 
\cite{stassis} is worthy to be considered. 
On the other hand, despite the soft mode the general features of the phonon
density of states till 60 meV are reasonably well described by the model 
\cite{dolgov}. 



There is an interesting similarity between the picture we arrived at above 
for MgB$_2$ and the picture based on the present knowledge on 
transition metal borocarbides such as YNi$_2$B$_2$C. The latter exhibit  
pronounced nesting properties of a rather complex Fermi surface (FS).
There the strongly coupled slow electrons stem from the nested 
parts of the FS. The anisotropy of the FS in borocarbides
is somewhat larger than in 
the present case but the anisotropy of the coupling in MgB$_2$ 
seems to be larger.
 In this context the $q$-dependent 
measurement of the soft phonon near 17 meV 
in the superconducting state is of 
great interest. Analogous behavior as for the anomalous 7 meV mode 
for YNi$_2$B$_2$C \cite{stassis} might be expected. 
The  $H_{c2}(0)$, the positive curvature of $H_{c2}(T)$ near $T_c$,
 and the sizeable
deviations of the shape 
of $H_{c2}(T)$ at low $T$ 
from the parabolic WHH (Werthamer-Helfand-Hohenberg)
-shape in both systems can be ascribed to 
the interband coupling with the remaining weakly coupled fast quasiparticles.

In conclusion, the proposed ISB criterion 
excludes the applicability of the frequently used isotropic 
single band model. The shape and the magnitude of the 
 upper critical field in MgB$_2$ can be successfully
described within a {\it multi}-band Eliashberg model with two
options. Provided the modern band structure calculations describe
correctly the electronic structure near the Fermi surface, then there
should be a sizeable coupling to some 
soft {\it bosonic} modes, not necessarily phonon-like,  
 not predicted by 
harmonic phonon
calculations but seen possibly in the neutron data. Otherwise the bare
quasiparticle Fermi velocities should be considerably smaller than the 
band structure predictions.

We thank O.\ Dolgov, M.\ Richter, W.\ Weber,  and T.\ Mishonov for 
discussions and J.\ Freudenberger 
for technical support.  This work was supported by the SFB 463 and the 
Deutsche Forschungsgemeinschaft.

\begin{table}
\caption{ ISB check for selected superconductors. 
Input quantities: $\omega_{pl}$ in eV,  $H_{c2}(0)$ in Tesla, 
cell volume $V$ in \AA$^3$, Sommerfeld constant   $\gamma_S$ in mJ/mol K$^2$,
T$_c$ in K, coupling constant $\lambda$, and the ISB parameter $Q$  (Eq.\ (1)).
Row 3: weak coupling, $\langle \Omega \rangle $ =60 meV; row 4: strong coupling
$\langle \Omega \rangle $=17.5 meV}
\begin{tabular}{c|c|c|c|c|c|c||c}
\ & $\omega_{pl}$  & $H_{c2}(0)$  & $V$ & $\gamma_S$  & $T_c$    & $\lambda$   & 
$Q$  \\ 
\tableline
Nb$\ ^a$            & 9.9 & 0.35 &  18   & 7.8 & 9.3    & 0.9  &  1.4 \\ 
YNi$_2$B$_2$C$\ ^b$   & 4.0 & 10   &  64   & 19  &  15    & 0.7  &  4.1  \\
MgB$_2\ ^c$         & 7.0 & 18   &  29   & 3.0 &  40    & 0.8  &  8.4   \\
MgB$_2\ ^d$         & 7.0 & 18   &  29   & 3.0 &  40    & 2.5  &  3.3   \\
\end{tabular}
{\small $^a$Ref.\onlinecite{weber91,lynch} and references therein.\\
$^b$Ref.\onlinecite{shulga98} and references therein.\\
$^c$Refs.\onlinecite{dolgov,canfield,budko}, weak coupling scenario,
 $\langle\Omega\rangle$=60 meV\\
$^d$Ibid, strong coupling scenario, $\langle\Omega\rangle$=17 meV\\}

\label{tab1}
\end{table}

\begin{figure}
\psfig{figure=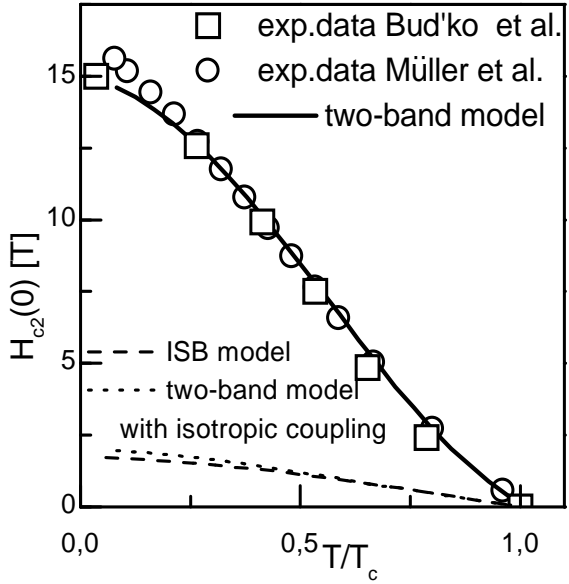,width=8.0cm,height=7.9cm}
\caption{ Experimental data for $H_{c2}(T)$ of
MgB$_2$ 
 compared with the theoretical curve of the two-band model for 
 both 
 anisotropic coupling and Fermi velocities (full line).
 Parameter set see text. For comparison the case of a complete isotropic 
 single band 
 model (ISB) (dashed) and the  two-v$_F$-band model with
  isotropic coupling 
 (dotted line) are shown, too.}
 \label{fig2}
 \end{figure}

\end{document}